\documentclass[conference]{IEEEtran}
\IEEEoverridecommandlockouts
\usepackage{amssymb}
\usepackage{amsthm}
\usepackage{multirow}
\usepackage{inputenc}
\usepackage{enumerate} 
\usepackage{textcomp}
\usepackage{listings}
\usepackage{array}
\usepackage[switch,mathlines]{lineno}
\usepackage{soul}
\usepackage{xfrac}
\usepackage{bbm}
\usepackage{breqn}
\usepackage{graphicx}
\usepackage{cite}
\usepackage{algorithm,algpseudocode}
\algnewcommand{\Inputs}[1]{%
  \State \textbf{Inputs:}
  \Statex \hspace*{\algorithmicindent}\parbox[t]{.8\linewidth}{\raggedright #1}
}
\algnewcommand{\Initialize}[1]{%
  \State \textbf{Initialize:}
  \Statex \hspace*{\algorithmicindent}\parbox[t]{.8\linewidth}{\raggedright #1}
}
\usepackage{algpseudocode}

\usepackage{cancel}
\usepackage{cite}
\usepackage{bm}
\usepackage{subfig}
\usepackage{float}
\def\BibTeX{{\rm B\kern-.05em{\sc i\kern-.025em b}\kern-.08em
    T\kern-.1667em\lower.7ex\hbox{E}\kern-.125emX}}

\begin{document}
\title{Identification of Power System Oscillation Modes using Blind Source Separation based on Copula Statistic\\
\thanks{The authors gratefully acknowledge the financial support of NSF via the grant ID 1917308.}
}
% \author{%
%   Pooja Algikar\IEEEauthorrefmark{1},
%   Lamine Mili\IEEEauthorrefmark{1},
%   Mohsen Ben Hassine \IEEEauthorrefmark{2},
%   Somayeh Yarahmadi\IEEEauthorrefmark{1}, \\
%   Almuatazbellah (Muataz) Boker \IEEEauthorrefmark{1},
%   \\
%   \IEEEauthorblockA{%
%     \IEEEauthorrefmark{1}Electrical and Computer Engineering, Northern Virginia Center, Virginia Tech, USA}\\
%   \IEEEauthorblockA{%
%     \IEEEauthorrefmark{2}Department of Computer Science, University of El Manar, Tunisia
%   }
% }
\author{\IEEEauthorblockN{Pooja Algikar}
\IEEEauthorblockA{\textit{Electrical and Computer Engineering}\\
\textit{Northern Virginia Center, Virginia Tech}\\
apooja19@vt.edu}\\
\and
\IEEEauthorblockN{Lamine Mili}
\IEEEauthorblockA{\textit{Electrical and Computer Engineering} \\
\textit{Northern Virginia Center, Virginia Tech}\\
lmili@vt.edu}\\
\and
\IEEEauthorblockN {Mohsen Ben Hassine}
\IEEEauthorblockA{\textit{Department of Computer Science} \\
\textit{University of El Manar, Tunisia}\\
mohsenmbh851@gmail.com}\\
\and
\IEEEauthorblockN{Somayeh Yarahmadi}
\IEEEauthorblockA{{\textit{Electrical and Computer Engineering}}\\
\textit{Northern Virginia Center, Virginia Tech}\\
syarahmadi@vt.edu}\\
\and
\IEEEauthorblockN{Almuatazbellah (Muataz) Boker}
\IEEEauthorblockA{{\textit{Electrical and Computer Engineering}}\\
\textit{Northern Virginia Center, Virginia Tech}\\
boker@vt.edu}\\
}
\maketitle
\begin{abstract}
The dynamics of a power system with large penetration of renewable energy resources are becoming more nonlinear due to the intermittence of these resources and the switching of their power electronic devices. Therefore, it is crucial to accurately identify the dynamical modes of oscillation of such a power system when it is subject to disturbances to initiate appropriate preventive or corrective control actions. In this paper, we propose a high-order blind source identification (HOBI) algorithm based on the copula statistic to address these non-linear dynamics in modal analysis. The method combined with Hilbert transform (HOBI-HT) and iteration procedure (HOBMI) can identify all the modes as well as the model order from the observation signals obtained from the number of channels as low as one. We access the performance of the proposed method on numerical simulation signals and recorded data from a simulation of time domain analysis on the classical 11-Bus 4-Machine test system. Our simulation results outperform the state-of-the-art method in accuracy and effectiveness.
\end{abstract}

\section{Introduction}
The increasing integration of renewable energy sources (RES) and distributed generators (DGs) accompanied by power electronic switching devices to the electric power system grid contribute majorly to generating signal oscillations \cite{Sivakumar2013StabilityPSS}. The caveat above is exacerbated when they result in an unbalance in the synchronicity of the generating sources causing errors in the fault current calculations, which lead to the mal-coordination of the protective equipment \cite{Saraf2012FaultGeneration}. Power system stabilizers are used in these scenarios to enhance system stability by damping out the unwanted oscillations whose parameter settings depend on the modal analysis of the underlying system. Therefore, it is crucial to accurately perform modal analysis addressing the non-linear dynamics in the RES and DG-integrated modern electric grid to identify all the modal frequencies. 

Multiple modal analysis methods presented in the literature can be categorized as model-based methods 
\cite{Lauria2014OnDetection},
\cite{Wang2018OnlineMatching} and data-driven methods,  \cite{Huang1998TheAnalysis,Kamwa2011RobustOperator,Laila2009AMonitoring,Papadopoulos2016Measurement-BasedSystems,Rueda2011Wavelet-basedOscillations,Senroy2007AnQuality,Tashman2014Multi-dimensionalSynchrophasors,Wadduwage2015IdentificationModels,Zhou2012AModes}, have been flourishing in recent years as phasor measurement units (PMUs) and wide area monitoring systems (WAMS) are deployed in power systems. Unlike model-based methods, data-driven methods do not require the knowledge of accurate models of ever-evolving real power systems. The methods are free from the complex matrix calculations associated with the growing transmission and distribution grids. Blind source separation (BSS) techniques are used to extract the latent, uncorrelated source signals from the observed multivariate time series data that is assumed to be a linear mixture of source signals.

The algorithms proposed for modal analysis using BSS include the second-order blind identification (SOBI) \cite{Belouchrani1997AStatistics} and its special case the algorithm for multiple unknown signal extraction (AMUSE) \cite{Tong1990AMUSE:Algorithm}. The latter makes use of the temporal dependence of the components via joint diagonalization of more or one time-shifted auto-correlation matrices, respectively \cite{Pan2021AMethods}. However, these separation methods consider only second-order dependencies and uncorrelation among source components while in some cases these components are non-Gaussian independent. Furthermore, they fail to consider high-order dependencies between the components, which increase with the unbalance and non-linear dynamics of the system. In other words, they do not account for the non-linear coupling between states in settings such as delay in fault clearing, thus yielding inaccurate modal analysis results.

To address these weaknesses, we propose in this paper a high-order blind modal identification algorithm (HOBMI) for power system dynamic stability analysis. The development of the HOBMI is threefold. Firstly, we initiate a high-order blind identification (HOBI) method based on the copula statistic (CoS), which accounts for the nonlinear dependency between the latent signals. We then combine the HOBI method with HT for estimating the modal parameters such as frequency and damping coefficient, yielding the HOBI-HT. Thirdly, we develop an iterative procedure for modal order identification of the HOBMI when the number of modes is unknown and when the observations are available in one channel. We demonstrate the proposed method on a numerical model and an 11-bus 4-machine test system. 

The remainder of this paper is organized as follows. Section II briefly introduces the copula and modal analysis methods of power systems using the blind source separation technique. Section III presents the proposed method. Section IV discusses the simulation results and Section V concludes the paper with a discussion of future work. 
\section{Background}
\subsection{A Brief Introduction on Copula}
Let us consider a $p$-dimensional random vector $(X_{1},\hdots,X_{p})$ with marginal cumulative distributions $F_{X_{1}}(X_{1}),\hdots,F_{X_{p}}(X_{p})$ respectively. A copula function ${C}_{\mathbf{X}}(\cdot)$: $[0,1]^{p}\rightarrow[0,1]$ 
is a unique function of marginal distribution yielding a joint cumulative probability distribution, thereby giving a dependency structure between the random variables $(X_{1},\hdots, X_{p})$ \cite{Sklar1959FonctionsMarges}.  The copula of $(X_{1},\hdots,X_{p})$ is formally given by 
\begin{equation}
\small
    C_{\mathbf{u}}=P(F_{X_1}(X_1)\leq u_{1},\hdots,F_{X_p}(X_p)\leq u_{p}),
\end{equation}
and the joint cumulative probability distribution as a function of copula $C_{\mathbf{X}}$  represented as 
\begin{equation}
\small
 F_{\bm{X}}(\bm{x})  = {C}_{\bm{X}}(F_{X_1}(x_1),F_{X_2}(x_2),\hdots,F_{X_p}(x_p)).
\end{equation}
The copula density of random vectors $X_{1},\hdots,X_{p}$ as 
\begin{equation}
\small
   \bm{c}_{\bm{X}}(F_{X_1}(x_1),\hdots,F_{X_p}(x_p))= \frac{\partial^{p} C_{\bm{X}}(F_{X_1}(x_1),\hdots,F_{X_p}(x_p)}{\partial F_{X_1}(x_1) \cdots F_{X_p}(x_p)}.
\end{equation}
Similarly, the joint probability density of random vectors, $f_{\bm{X}(\bm{x})}$ with marginal probability densities $f_{X_{1}}(x_{1}),\hdots, f_{X_{p}}(x_{p})$ is given by 
\begin{equation}
   f_{\bm{X}(\bm{x})}=\prod_{i=1}^{p} f_{X_i}(x_i)c_{\bm{X}}\left(F_{X_1}(x_1),\hdots,F_{X_p}(x_p)\right) 
\end{equation}

\subsection{Power Swing Analysis}
A measured electromechanical oscillation signal in a power system following a short-circuit can be regarded as a mixture of either local modes  involving single generators oscillating against the grid, inter-area modes involving groups of generators oscillating against each other, or a coherent swing oscillation of a group of generators against the remaining generators \cite{Raak2016Data-DrivenAnalysis}. This measured signal can be expressed as a sum of exponentially sinusoidal signals, termed modes. Formally, we have
\begin{equation}
\small
    x(t)=\sum_{j=1}^{N}A_{j}e^{\sigma_{j}t}sin(2\pi f_{j}t+\theta_{j}),
\end{equation}
where $A_{j},\sigma_{j},f_{j},\textrm{and}\; \theta_{j}$ are the relative amplitude, damping coefficient, oscillation frequency, and phase shift of the $j^{th}$ mode in $x(t)$, respectively in a system of $N$ different modes. 

Let us consider the amplitude modulated source signals corresponding to $N$ modes, $\bm{s}_{i}\in \mathbb{R}^N$, and the observed signals, $\mathbf{x}_{i}\in \mathbb{R}^q$, from $q$ channels measured at $T$ instances, $i=1,\hdots,T$. Let $\mathbf{A}\in \mathbb{R}^{N\times q}$ denote a mixing matrix of amplitude modulated source signals $\mathbf{S}=[\mathbf{s}_{1},\hdots,\mathbf{s}_{T}]$. Formally, the observed signals $\mathbf{X}=[\mathbf{x}_{1},\hdots,\mathbf{x}_{T}]$ are linearly related to the source signals, yielding
\begin{equation}
\small
    \mathbf{X}= \mathbf{A}\mathbf{S}.
\end{equation}  
In practice, we have access to the observed signals $\mathbf{X}$. Here, the problem is to infer the parameters of $\mathbf{S}$ from the estimated source signals $\mathbf{Y}$. To solve this problem, we may apply the classical blind source separation of linearly mixed signals that may not be non-Gaussian dependent to obtain an approximate estimate of source signals through the determination of the de-mixing matrix, $\mathbf{W}$. Formally we have
\begin{equation}\label{1}
\small
    \mathbf{Y}=\mathbf{W}\mathbf{X}.
\end{equation}
% SOBI, AMUSE, and FastICA are the algorithms proposed in the literature for the blind source separation (BSS) of the observation matrix $\mathbf{X}$. 

To estimate the de-mixing matrix, the SOBI algorithm may be used.  It is based on the joint diagonalization of time-shift covariance matrices of the whitened measured signals. Another algorithm similar to SOBI is AMUSE, except that for the latter the diagonalization of a single covariance matrix without time-shifting is employed \cite{Pan2021AMethods}. Techniques such as Hilbert transform (HT) are elicited to infer instantaneous modal parameters, including frequencies,  amplitudes, and damping coefficients \cite{Zhang2017IdentificationSeparation}.

\section{The Proposed Method}
We aim to identify the system modes using a high-order dependency matrix between the observed signals based on the copula statistic (CoS) developed in \cite{Ben2017ALearning}.
The resulting dependency matrix is jointly diagonalized using the popular joint approximation method (JAD) \cite{McNeill2008ADiagonalization}. We then propose a modified iteration algorithm for model order identification similar to \cite{Zhang2017IdentificationSeparation}. We assume that a required number of observations obtained from observation channels corresponding to each synchronous generator in the power system are accessible.
\subsection{Dependency Matrix Estimation Using the Copula Statistic}
The CoS is a statistical index that measures the strength of bi-variate linear and nonlinear dependence  \cite{Ben2017ALearning}. The corresponding measure relies on the properties such as concordance, quadrant dependence, and comonotonicity introduced by Lehmann \cite{Statistics1966SomeHttps://www.jstor.org/stable/} between any two random variables in terms of a relative distance function between the empirical copula, the Fréchet-Hoeffding bounds, and the independence copula. The CoS reduces to  Pearson’s correlation coefficient for the Gaussian copula and Spearman’s correlation coefficient for many copulas when using large sample sizes. It ranges from zero to unity, attaining its lower and upper limit for the case of independence and functional dependence, respectively. The steps for the estimation of CoS between two random variables $X$ and $Y$ with sample size $n$ are described as follows:
\begin{itemize}
\item  \textit{Step 1:} Calculate $u_{j}$, $v_{j}$ and $C_{n}(u,v)$ as follows: 
\begin{itemize}
    \item $u_{j}=\frac{1}{n}\sum_{j=1}^{n}\mathbbm{1}(k\neq j:x_{k}\leq x_{j})$
    \item $v_{j}=\frac{1}{n}\sum_{j=1}^{n}\mathbbm{1}(k\neq j:y_{k}\leq y_{j})$
    \item $C_{n}(u,v)=\frac{1}{n}\sum_{j=1}^{n}\mathbbm{1}(u_{j}\leq u, v_{j}\leq v)$
\end{itemize}
\item \textit{Step 2:} Order $x_{j}^{'}$s to get $x(1)\leq \hdots \leq x(n)$, which results in  $u(1)\leq \hdots \leq u(n)$ since $u_{j}=\frac{R_{x_{j}}}{n}$, where $R_{x_{j}}$ is the rank of $x_{j}$;
\item  \textit{Step 3:} Determine the domains $\mathcal{D}_{i};i=1,\hdots,m,$ where each $\mathcal{D}_{i}$ is a u-interval associated with a non-decreasing or non-increasing sequence of $C_{n}(u_{j},v_{p}), j=1,\hdots,n$; 
\item  \textit{Step 4:} Determine the smallest and the largest value of $C_{n}(u,v)$, denoted by $C_{i}^{min}$ and $C_{i}^{max}$, and find the associated $u_{i}^{min}$ and $u_{i}^{max}$ for each domain $\mathcal{D}_{i};i=1,\hdots,m,$;
\item \textit{Step 5:} Calculate $\lambda(C_{i}^{min})$ and $\lambda(C_{i}^{max})$ as 
\begin{equation*}
\small
    \lambda(C(u,v))=\begin{cases}
    \frac{C(u,v)-uv}{Min(u,v)-uv},&  \textrm{if}\; C(u,v)\geq uv,\\
    \frac{C(u,v)-uv}{Max(u+v-1,0)-uv}, & \textrm{if}\; C(u,v)< uv;
\end{cases}
\end{equation*}
\item \textit{Step 6:} If $\lambda(C_{i}^{min})$ and $\lambda(C_{i}^{max})$ are equal to one, go to Step 8; 
\item \textit{Step 7:} Calculate the absolute difference between the three consecutive values of $C_{n}(u_{(i)},v_{j})$ centered at $u_{i}^{min}$ (respectively at $u_{i}^{max}$) and decide that the central point is a local optimum if (i) both absolute differences are smaller than or equal to $\frac{1}{n}$ and (ii) there are more than four points within the two adjacent domains, $\mathcal{D}_{i}$ and $\mathcal{D}_{i+1}$;
\item \textit{Step 8:} Calculate $\gamma_{i}$ given by 
\begin{equation*}
\small
   \gamma_{i}=\begin{cases}
    1,&  \textrm{at local optimum of}\; Y=f(X)\; \textrm{on}\; D_{i},\\
    \frac{\lambda(C_{i}^{min})+\lambda(C_{i}^{max})}{2}, & \textrm{otherwise;}
\end{cases}.
\end{equation*}
\item \textit{Step 9:} Repeat Steps 2 through 7 for all the m domains, $\mathcal{D}_{i};i=1,\hdots,m,$;
\item Calculate the CoS given by 
$
    \textrm{CoS}(X,Y)=\frac{1}{n+m-1}\sum_{i=1}^{m}n_{i}\gamma_{i}.$
\end{itemize}
The dependency matrix $\mathbf{D}$ based on CoS is now estimated through 
\begin{equation}\label{2}
\small
    D_{i,j}=\textrm{CoS}(\mathbf{x}_{i},\mathbf{x}_{j}), \forall\; i,j=1,\hdots,q.
\end{equation} 
Next, we further enhance the existing SOBI algorithm to get the high-order blind identification algorithm (HOBI).
\subsection{High-Order Blind Identification (HOBI)}
The proposed HOBI method based on the dependency matrix, $\mathbf{D}$, consists of the following steps:  
 \begin{itemize}
     \item \textit{Step 1:} Estimate the dependency matrix $\mathbf{D}_{\mathbf{X}}(0)$ using \eqref{2} and perform an eigenvalue decomposition to get  $\mathbf{D}_{\mathbf{X}}(0)=\mathbf{P}\mathbf{\Lambda}\mathbf{P}^{T}$. 
     \item \textit{Step 2:}
     Obtain a set of statistically independent signals in $Z$ from $X$ as follows: $\mathbf{Z}=\mathbf{W}_{X}\mathbf{X}$, where $\mathbf{W}_{X}=\mathbf{\Lambda}^{\frac{-1}{2}}\mathbf{P}^{T}\mathbf{X}$.
     \item \textit{Step 3:} Calculate a set of time-delayed dependency matrices ${D}_{{Z}_{i,j}}(\tau)=\textrm{CoS}\left(\mathbf{z}_{i}(\tau:T),\mathbf{z}_{j}(1:T-\tau)\right);\; i,j=1,\hdots,q$ of $\mathbf{Z}$, where $\tau \in \{\tau_{a}|a=1,\hdots,n^{*}\}$ and $n^{*}=\textrm{min}(100, [\frac{T}{3}+1])$. 
     \item \textit{Step 4:} Compute the orthogonal normalized matrix $\psi^{T}$ by jointly diagonalizing the set $\{\mathbf{D}_{Z}(\tau_{a})|a=1,\hdots,n\}$ using JAD . 
     \item \textit{Step 5:} Calculate the de-mixing matrix $\mathbf{W}=\psi^{T}\mathbf{W}_{X}$ and evaluate the estimate of separated source signals $\mathbf{Y}=\mathbf{W}\mathbf{X}$
 \end{itemize}

In summary, the proposed HOBI algorithm allows us to  estimate the de-mixing matrix $\mathbf{W}$ in \eqref{1} to obtain the source estimates, $\mathbf{Y}$. To extract instantaneous modal properties of the estimated signals, $\mathbf{Y}$, Hilbert transform (HT) technique is employed.  Till now, we have not addressed the problem of model order identification. The HOBI-HT method can be directly employed if the model order, $M$, is known a priori and the received number of observations $q$ is greater than $M$; otherwise, it should be estimated as described next.  

\subsection{High-order Blind Modal Identification Algorithm (HOBMI)}
Modal analysis based on HOBI-HT is combined with the modified iteration algorithm to identify model order $M$, which involves the iteration over the assumed model order $m$ from 2 to $M_{max}$. In scenarios where $M>q$, Takens' embedding theory \cite{Yap2011StableSystems} is employed to constitute the number of required observations $q=M$.  

The proposed HOBMI method consists of the following steps: 
\begin{itemize}
    \item \textit{Step 1:} Identify the data window $L$ involving the oscillations predominantly to acquire adequate information from the recorded real-time observed signals. 
    
    \item \textit{Step 2:} Let m=2. Constitute the observation matrix $\mathbf{X}$ from Takens' embedding theory with embedding dimensions $d^{*}=2m$.
    
    \item \textit{Step 3:} Apply the modal analysis method HOBI on the selected data window to obtain estimates of sources $\mathbf{Y}$.
     Then, apply the Hilbert transform technique to obtain the instantaneous frequency $f_{i}$ and damping $\sigma_{i}$, for $i=1,\hdots,2m$ of the corresponding estimated signal from $\mathbf{Y}=[\mathbf{y}_{1},\hdots,\mathbf{y}_{2m}]$. 
    
    \item \textit{Step 4:} Sort the obtained frequencies $\bm{f}$ and calculate the difference between adjacent elements. Sort again the elements of $\bm{f}$ in ascending order of adjacent difference. Find the corresponding damping component from $\bm{\sigma}$. The two entries $f_{k1},f_{k2}$ with minimal difference in $\mathbf{f}$ is corresponding to complex conjugate pair of the $k^{th}$ mode with $m$. As for the damping coefficient, it is the corresponding component in $\bm{\sigma}$, $\sigma_{k1},\sigma_{k2}$.  
    
    \item \textit{Step 5:} Average the frequencies and damping coefficients in the $k^{th}$ mode and gather them into ${F}_{avg}$ and ${\Sigma}_{avg}$ as ${F}_{avg}(m-1,k)=\frac{f_{kk}+f_{kk+1}}{2}$ and ${\Sigma}_{avg}(m-1,k)=\frac{\sigma_{kk}+\sigma_{kk+1}}{2}$, $\;k=1,\hdots,m$; $kk=1,\hdots,2m-1$. To calculate the divergence between the two modes in $m$, we use the divergence index ${D}_{div}(m-1,k)=\sqrt{(f_{kk}^{2}-f_{kk+1}^{2})+(\sigma_{kk}^{2}-\sigma_{kk+1}^{2})}$.
    
    \item \textit{Step 6:} Let $m=m+1$ and iterate the steps from Step 3 to Step 5 until $m>M_{max}$ to obtain the matrices $\mathbf{F}_{avg}, \mathbf{\Sigma}_{avg}$, and $\mathbf{D}_{dvg}$.
    \item \textit{Step 7:} 
    Find the minimum value in the divergence matrix $\mathbf{D}_{dvg}$ in each column and get the corresponding frequency and damping from the average frequency matrix $\mathbf{F}_{avg}$ and the average damping coefficient matrix $\mathbf{\Sigma}_{avg}$.   
\end{itemize}
Interestingly, using the HOBMI, the modal parameters of the underlying system can be determined even under circumstances where the protective systems fail to respond in a timely manner. The proposed HOBMI algorithm outlined above is summarized in Algorithm \ref{alg1}.  Next, we assess the performance of the HOBMI.
\begin{algorithm}[!htbp]
\small
\caption{High Order Blind Modal Identification (HOBMI)}\label{alg1}
\begin{algorithmic}
\State \textbullet\ Obtain the data $\mathbf{x}$ within $L$ size window
\State \textbullet\ Initialize k=1,kk=1
\For {$m:1:M_{max}$}
\State {$\boldsymbol{-}$} Build the observation matrix $\mathbf{X}$ from Takens' embedding theory with $d^{*}=2m$
\State {$\boldsymbol{-}$} Perform HOBI-HT based modal analysis
\State {$\boldsymbol{-}$} Obtain the set of instantaneous frequencies $\bm{f}=[f_{1},\hdots,f_{2m}]$ and $\bm{\sigma}=[\sigma_{1},\hdots,\sigma_{2m}]$
\State {$\boldsymbol{-}$} Sort $\bm{f}$ and find corresponding rearranged $\bm{\sigma}$
\If{$2j \leq length(\bm{f})$}
    \State {$\boldsymbol{-}$} $F_{avg}\left(m-1,j\right)=\frac{f(jj)+f(jj+1)}{2}$ 
    \State {$\boldsymbol{-}$} $\Sigma_{avg}\left(m-1,j \right)=\frac{\sigma_{kk}+\sigma_{kk+1}}{2}$
    \State {$\boldsymbol{-}$} $D_{div}\left(m-1,j\right)=\sqrt{(\sigma(kk)-\sigma(kk+1))^2 +(f\left(kk\right)-f\left(kk+1\right))^2}$
    \State {$\boldsymbol{-}$} $kk = kk+2, k = k+1 $
\EndIf
\EndFor 
\State \textbullet\ Get $\mathbf{F}_{avg}$, $\mathbf{\Sigma}_{avg}$ , $\mathbf{D}_{dvg}$
\State \textbullet\ [a,b] = size$(\mathbf{D}_{div})$, $ii =1$ 
\If{$(ii \leq b)$ and length(find ($\mathbf{D}_{div} \left(:,ii == NaN \right)$) $<$ 4)}
    \State {$\boldsymbol{-}$} $\left[m_{val},m_{det}\right]=\min (\mathbf{D}_{div} \left(:,ii \right))$\;
    \State {$\boldsymbol{-}$} $f_{det} \left(1,ii \right) = f_{av} \left(m_{det}, ii \right)$
    \State {$\boldsymbol{-}$} $\sigma_{det}(1,ii)={\Sigma}_{av}(m_{det},ii) $
    \State {$\boldsymbol{-}$} $ii = ii + 1$  
\EndIf
\end{algorithmic}
\end{algorithm}
\section{Performance Evaluation of the HOBMI Method}
Simulations are carried out to compare the performance of the proposed HOBMI to that of the SHDMI proposed in \cite{Zhang2017IdentificationSeparation} given their  similarities.  We demonstrate the methods of modal analysis on a numerical system and on an 11-bus 4-machine test system.  
\subsection{Numerical Model}
We generate two synthetic signals given by 
\begin{equation}
\small
 x_{1}(t)=\begin{cases}
     1e^{-0.01}cos(8t) &0\leq t\leq2, 6\leq t\leq10\\
      0 & 2\leq t\leq6\\
 \end{cases}
 \end{equation}
 \begin{equation}
x_{2}(t)= \begin{cases}
     x_{2}(t)=0 & 0\leq t\leq6\\
      0.6e^{-0.03}cos(17t) & 6\leq t\leq10\\
 \end{cases} 
 \end{equation}
To introduce dependency among the two time-series signals $x_{1}(t)$ and $x_{2}(t)$, two random sequences, $w_{1}$ and $w_{2}$, are drawn from a bi-variate probability distribution associated with each of the Archimedian copula family, namely, Frank, Clayton, Gumbel, and Gaussian copula with a specified parameter ${\alpha}$. They are then added to $\mathbf{x}_{1}$ and $\mathbf{x}_{2}$, respectively to form two source signals, $s_{1}(t)=x_{1}(t)+w_{1}(t)$ and $s_{2}(t)=x_{2}(t)+w_{2}(t)$. 
The observation signal is expressed as $x(t)=s_{1}(t)+s_{2}(t)$, which contains two modes. Then, both the SOBI-HT and HOBI-HT are applied to estimate the frequencies and damping coefficients of the source signals $\mathbf{s}_{1}$ and $\mathbf{s}_{2}$. 
Note that, the frequency and damping coefficients obtained from the HT method applied to $\mathbf{s}_{1}$ and $\mathbf{s}_{2}$ now stand as unambiguous modal estimates for accuracy assessment of the results obtained from SOBI-HT and HOBI-HT. 
\begin{table}[]
\small
\setlength{\tabcolsep}{3.6pt}
\centering
\caption{Results for the Numerical model}
  \begin{tabular}{|c|cc|cccc|}
\hline
Mode & \multicolumn{2}{|c|}{HT of $\mathbf{s}_{1},\mathbf{s}_{2}$} &\multicolumn{2}{c|}{HOBI-HT} &\multicolumn{2}{c|}{SOBI-HT} \\
\hline
{} & ${f_{s}}$ & $\sigma_{s}$ & ${f_{i}}$ & $\sigma_{i}$ & ${f_{i}}$ & $\sigma_{i}$\\
\hline
{} & {}&{} & \multicolumn{4}{c|}{$w_{1},w_{2}\sim$ Frank$(\alpha=10)$}\\
\hline
1&0.78972&0.26874&0.77953&0.26235&0.79294&0.26286\\
2&1.2082&0.65909&1.3193&0.58809&2.097&0.68111\\\hline
{} & {}&{} & \multicolumn{4}{c|}{$w_{1},w_{2}\sim$ Clayton$(\alpha=10)$}\\
\hline
1&0.78957&0.26854&0.77578&0.26228&0.79271&0.26282\\
2&1.2081&0.65807&1.4304&0.57673&2.6525&0.68365\\
\hline
{} & {} &{}& \multicolumn{4}{c|}{$w_{1},w_{2}\sim$ Gumbel$(\alpha=10)$}\\
\hline
1&0.78887&0.26872&0.77134&0.26241&0.79196&0.26285\\
2&1.2081&0.65508&1.3193&0.55595&1.8747&0.67585\\\hline
{} & {} &{} & \multicolumn{4}{c|}{$w_{1},w_{2}\sim$ Gaussian$(\alpha=0.9)$}\\
\hline
1&0.78919&0.26807&0.7753&0.2616&0.79249&0.26221\\
2&1.208&0.66199&1.3192&0.5823&2.0969&0.68541\\\hline
\end{tabular}
\label{tabnum}
\end{table}

The results are displayed in Table \ref{tabnum}.
We observe that the frequency of the second source signal for all the cases of copula estimates much closer to the true frequency of $1.0282$ in the results obtained from HOBI-HT than SOBI-HT. Even in the case of a Gaussian correlation of $0.9$, SOBI-HT yields unreliable estimates. On the other hand, the HOBI-HT method yields better estimations of the modal parameters for all the cases of the considered copula family.  
\subsection{Simulations of the 11-Bus 4-Machine Test System}
\begin{figure}%
    \centering
  \subfloat[\centering ]{{\includegraphics[height=2.1cm,width=8.2cm]{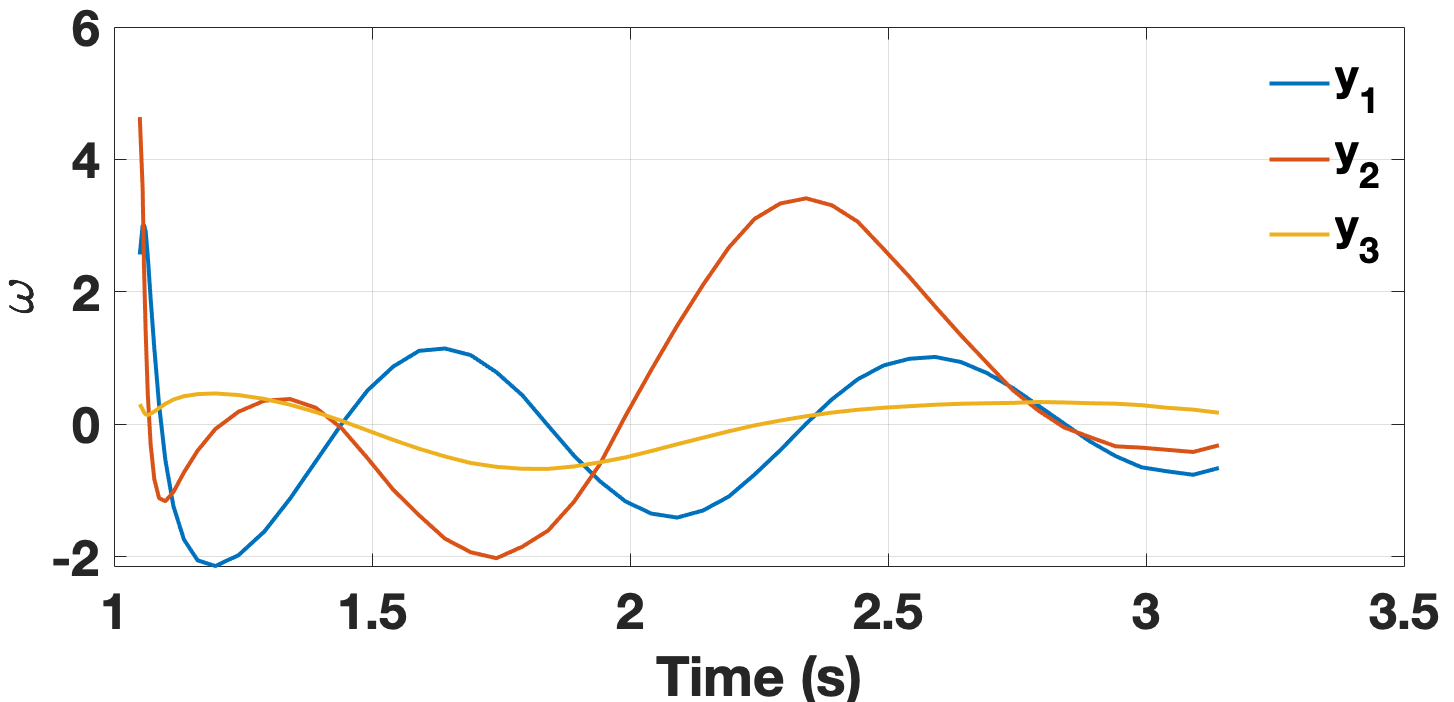}}}%
 \qquad 
    \subfloat[\centering ]{{\includegraphics[height=2.1cm,width=8cm]{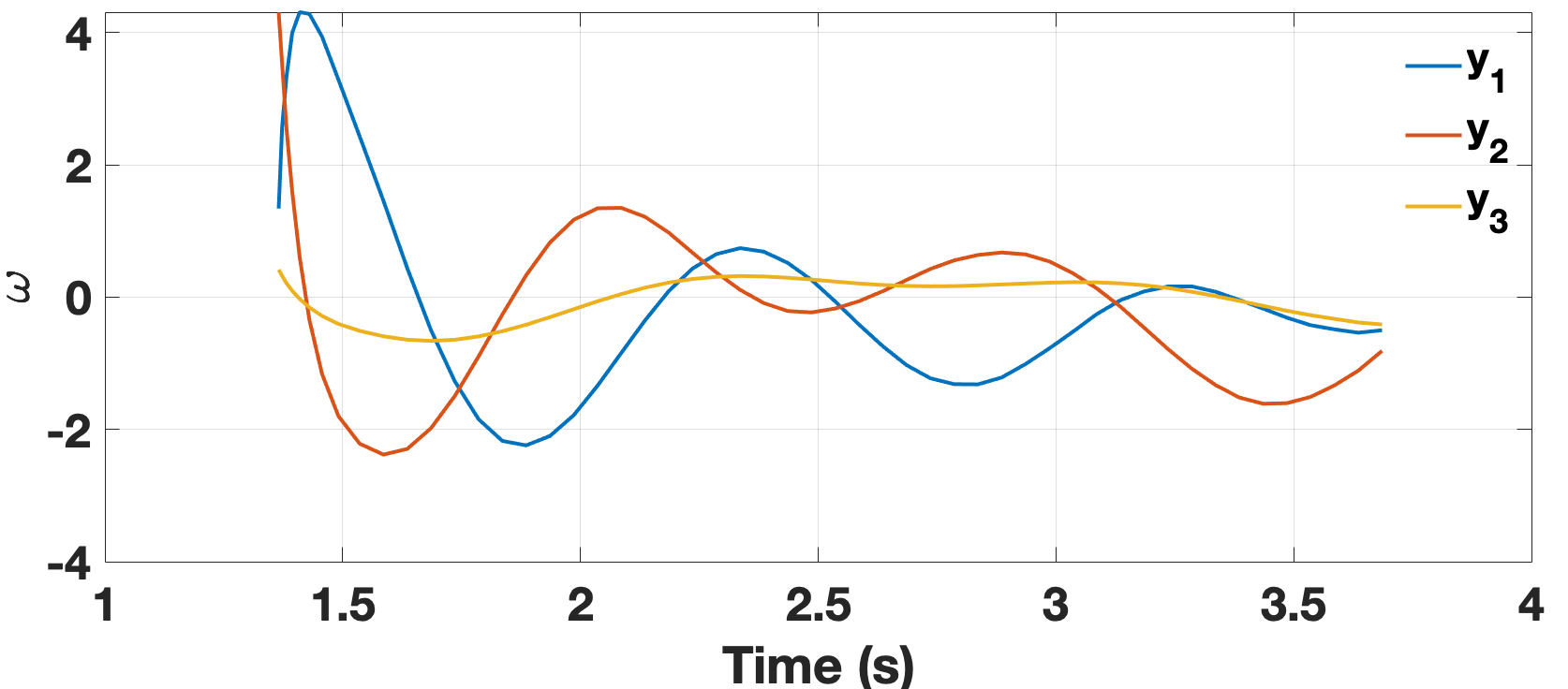}}}%
    \caption{{The separated sources from the HOBMI for (a) Case A; (b) Case B.}}%
    \label{r1}
\end{figure}

We consider the following two simulation cases of a time domain analysis with a three-phase unbalanced fault applied at bus 8:
(i) Case A: The fault is applied at $1$s and cleared at $1.05$s; (ii) Case B: The fault is applied at 1s and cleared at $1.3$s.
The simulations are performed in Matlab toolbox PSAT carried out on the classical $11$-bus $4$-machine test system \cite{kundur2022power}. The specified sampling frequency and the base frequency are  $20$ Hz and  $60$ Hz, respectively.
The results obtained from the linear eigenvalue analysis (LEA) are considered reference values of the modal parameters of the test system to access the accuracy of both methods. The most associated states (MAS) are found to be the angular velocities ($\omega_{syn}$) of  the synchronous machines 1,2, and 4 from LEA. Therefore, the relative angular velocity measurements $\omega_{24}\; \textrm{and} \;\omega_{14}$ serve as observed signals from two channels for both cases.
Table \ref{d_kundur1_mdl_01_05} and \ref{d_kundur1_mdl_01_1} for Cases A and B, respectively. We show that in Case B, where the non-linear coupling between states is dominant, our HOBMI method outperforms the SHDMI method. The separated signals obtained from the HOBMI method are displayed in Fig. \ref{r1}. The dataset used to obtain the experimental results and code is provided at https://github.com/apooja1/HOBMI. 
\begin{table}[t]
\small
\setlength{\tabcolsep}{0.5pt}
\centering
\caption{Case A Results for the 11-Bus 4-Machine System}
  \begin{tabular}{|c|c|cc|cc|cc|}
\hline
Eigenvalue & MAS & \multicolumn{2}{|c|}{LEA} & \multicolumn{2}{c}{HOBMI} & \multicolumn{2}{|c|}{SHDMI}\\
\hline
{Eig}  & {$\omega_{syn}$} & $f_{i}$ & $\sigma_{i}$ & $f_{i}$ & $\sigma_{i}$   & $f_{i}$ & $\sigma_{i}$\\
\hline
$\{ 9,10\}$  & $\omega_{{syn}{_1}}$ & $1.08368$ & $-0.56646$ & $1.0937$ & $-0.5057$   & $1.0743$ & $-0.5721$\\
 $\{11,12\}$  & $\omega_{{syn}{_2}}$ & $1.05323$ & $-0.55530$ & $1.0393$ & $ -0.5468$   & $ 1.0489$   & $-0.6033$\\
$\{13,14\}$  & $\omega_{{syn}{_4}}$ & $0.54358$ & $-0.13084$ & $  0.5443$ & $ -0.1284$   & $0.5499$ & $-0.1452$\\
\hline
\end{tabular}
\label{d_kundur1_mdl_01_05}
\end{table}

\begin{table}[t]
\small
\setlength{\tabcolsep}{0.5pt}
\centering
\caption{Case B Results for the 11-Bus 4-Machine System}
  \begin{tabular}{|c|c|cc|cc|cc|}
\hline
Eigenvalue & MAS & \multicolumn{2}{|c|}{LEA} & \multicolumn{2}{c}{HOBMI} & \multicolumn{2}{|c|}{SHDMI}\\
\hline
{Eig}  & {$\omega_{syn}$}& $f_{i}$ & $\sigma_{i}$ & $f_{i}$ & $\sigma_{i}$   & $f_{i}$ & $\sigma_{i}$\\
\hline
$\{9,10\}$  & $\omega_{{syn}{_1}}$ & $1.0851$ & $-0.56655$ & $1.0931$ &  $-0.3608$  &$1.6305$  &  $0.0027$\\
$\{11,12\}$  & $\omega_{{syn}{_2}}$ & $  1.0544 $ & $-0.55725 $ & $ 1.0571 $ &  $-0.3401$ & $1.1324$ & $-0.0660 $\\
 $\{13,14\}$ & $\omega_{{syn}{_4}}$ & $0.54585$ & $-0.12945$ & $  0.5302$ & $-0.1527$   & $1.4939$ & $-0.0083$\\
\hline
\end{tabular}
\label{d_kundur1_mdl_01_1}
\end{table}
\section{Conclusion and Future Work}
In this paper, a new blind source separation method based on the copula statistic has been developed. Simulations have highlighted the excellent performance of the proposed method when the dependencies of the source signals are nonlinear. As future work, we will robustify the proposed method against outliers and we will extend it to identify groups of generators that oscillate in an anti-phase motion. 
\bibliography{references2} 

\begin{thebibliography}{10}

\bibitem{Sivakumar2013StabilityPSS}
P.~Sivakumar and C.~Birindha, ``{Stability enhancement of DG sourced power
  system with modified AVR and PSS},'' {\em Proceedings of International
  Conference on Computation of Power, Energy, Information and Communication,
  ICCPEIC 2013}, pp.~105--109, 2013.

\bibitem{Saraf2012FaultGeneration}
P.~Saraf, ``{Fault Analysis of an Unbalanced Distribution System With
  Distributed Generation},'' no.~August, 2012.

\bibitem{Lauria2014OnDetection}
D.~Lauria and C.~Pisani, ``{On Hilbert transform methods for low frequency
  oscillations detection},'' {\em IET Generation, Transmission and
  Distribution}, vol.~8, no.~6, pp.~1061--1074, 2014.

\bibitem{Wang2018OnlineMatching}
C.~Wang, C.~Li, G.~Wu, G.~Li, and Z.~Du, ``{Online identification of power
  system oscillation modes based on mode shape matching},'' {\em China
  International Conference on Electricity Distribution, CICED},
  no.~201805160000003, pp.~1576--1580, 2018.

\bibitem{Huang1998TheAnalysis}
N.~E. Huang, Z.~Shen, S.~R. Long, M.~C. Wu, H.~H. Snin, Q.~Zheng, N.~C. Yen,
  C.~C. Tung, and H.~H. Liu, ``{The empirical mode decomposition and the Hubert
  spectrum for nonlinear and non-stationary time series analysis},'' {\em
  Proceedings of the Royal Society A: Mathematical, Physical and Engineering
  Sciences}, vol.~454, no.~1971, pp.~903--995, 1998.

\bibitem{Kamwa2011RobustOperator}
I.~Kamwa, A.~K. Pradhan, and G.~Joos, ``{Robust detection and analysis of power
  system oscillations using the Teager-Kaiser energy operator},'' {\em IEEE
  Transactions on Power Systems}, vol.~26, no.~1, pp.~323--333, 2011.

\bibitem{Laila2009AMonitoring}
D.~S. Laila, A.~R. Messina, and B.~C. Pal, ``{A refined Hilbert-Huang transform
  with applications to interarea oscillation monitoring},'' {\em IEEE
  Transactions on Power Systems}, vol.~24, no.~2, pp.~610--620, 2009.

\bibitem{Papadopoulos2016Measurement-BasedSystems}
T.~A. Papadopoulos, A.~I. Chrysochos, E.~O. Kontis, P.~N. Papadopoulos, and
  G.~K. Papagiannis, ``{Measurement-Based Hybrid Approach for Ringdown Analysis
  of Power Systems},'' {\em IEEE Transactions on Power Systems}, vol.~31,
  no.~6, pp.~4435--4446, 2016.

\bibitem{Rueda2011Wavelet-basedOscillations}
J.~L. Rueda, C.~A. Ju{\'{a}}rez, and I.~Erlich, ``{Wavelet-based analysis of
  power system low-frequency electromechanical oscillations},'' {\em IEEE
  Transactions on Power Systems}, vol.~26, no.~3, pp.~1733--1743, 2011.

\bibitem{Senroy2007AnQuality}
N.~Senroy, S.~Suryanarayanan, and P.~F. Ribeiro, ``{An improved Hilbert-Huang
  method for analysis of time-varying waveforms in power quality},'' {\em IEEE
  Transactions on Power Systems}, vol.~22, no.~4, pp.~1843--1850, 2007.

\bibitem{Tashman2014Multi-dimensionalSynchrophasors}
Z.~Tashman, H.~Khalilinia, and V.~Venkatasubramanian, ``{Multi-dimensional
  fourier ringdown analysis for power systems using synchrophasors},'' {\em
  IEEE Transactions on Power Systems}, vol.~29, no.~2, pp.~731--741, 2014.

\bibitem{Wadduwage2015IdentificationModels}
D.~P. Wadduwage, U.~D. Annakkage, and K.~Narendra, ``{Identification of
  dominant low-frequency modes in ring-down oscillations using multiple Prony
  models},'' {\em IET Generation, Transmission and Distribution}, vol.~9,
  no.~15, pp.~2206--2214, 2015.

\bibitem{Zhou2012AModes}
N.~Zhou, J.~W. Pierre, and D.~Trudnowski, ``{A stepwise regression method for
  estimating dominant electromechanical modes},'' {\em IEEE Transactions on
  Power Systems}, vol.~27, no.~2, pp.~1051--1059, 2012.

\bibitem{Belouchrani1997AStatistics}
A.~Belouchrani, K.~Abed-Meraim, J.~F. Cardoso, and E.~Moulines, ``{A blind
  source separation technique using second-order statistics},'' {\em IEEE
  Transactions on Signal Processing}, vol.~45, no.~2, pp.~434--444, 1997.

\bibitem{Tong1990AMUSE:Algorithm}
L.~Tong, V.~C. Soon, Y.~F. Huang, and R.~Liu, ``{AMUSE: A new blind
  identification algorithm},'' {\em Proceedings - IEEE International Symposium
  on Circuits and Systems}, vol.~3, pp.~1784--1787, 1990.

\bibitem{Pan2021AMethods}
Y.~Pan, M.~Matilainen, S.~Taskinen, and K.~Nordhausen, ``{A review of
  second-order blind identification methods},'' {\em Wiley Interdisciplinary
  Reviews: Computational Statistics}, no.~January, pp.~1--23, 2021.

\bibitem{Sklar1959FonctionsMarges}
M.~J. Sklar, ``{Fonctions de repartition a n dimensions et leurs marges},''
  1959.

\bibitem{Raak2016Data-DrivenAnalysis}
F.~Raak, Y.~Susuki, and T.~Hikihara, ``{Data-Driven Partitioning of Power
  Networks Via Koopman Mode Analysis},'' {\em IEEE Transactions on Power
  Systems}, vol.~31, no.~4, pp.~2799--2808, 2016.

\bibitem{Zhang2017IdentificationSeparation}
A.~Q. Zhang, L.~L. Zhang, M.~S. Li, and Q.~H. Wu, ``{Identification of Dominant
  Low Frequency Oscillation Modes Based on Blind Source Separation},'' {\em
  IEEE Transactions on Power Systems}, vol.~32, no.~6, pp.~4774--4782, 2017.

\bibitem{Ben2017ALearning}
M.~Ben, L.~Mili, and K.~Karra, ``{A Copula Statistic for Measuring Nonlinear
  Dependence with Application to Feature Selection in Machine Learning},'' {\em
  International Journal of Advanced Computer Science and Applications}, vol.~8,
  no.~7, pp.~144--154, 2017.

\bibitem{McNeill2008ADiagonalization}
S.~I. McNeill and D.~C. Zimmerman, ``{A framework for blind modal
  identification using joint approximate diagonalization},'' {\em Mechanical
  Systems and Signal Processing}, vol.~22, no.~7, pp.~1526--1548, 2008.

\bibitem{Statistics1966SomeHttps://www.jstor.org/stable/}
M.~Statistics, ``{Some Concepts of Dependence Author ( s ): E . L . Lehmann
  Source : The Annals of Mathematical Statistics , Oct ., 1966 , Vol . 37 , No
  . 5 ( Oct ., 1966 ), pp . Published by : Institute of Mathematical Statistics
  Stable URL : https://www.jstor.org/stable/},'' vol.~37, no.~5,
  pp.~1137--1153, 1966.

\bibitem{Yap2011StableSystems}
H.~L. Yap and C.~J. Rozell, ``{Stable takens' embeddings for linear dynamical
  systems},'' {\em IEEE Transactions on Signal Processing}, vol.~59, no.~10,
  pp.~4781--4794, 2011.

\bibitem{kundur2022power}
P.~S. Kundur and O.~P. Malik, {\em {Power system stability and control}}.
\newblock McGraw-Hill Education, 2022.

\end{thebibliography}
\bibliographystyle{ieeetr}

\end{document}